\documentclass[aps,twocolumn,superscriptaddress,prl,fleqn,showpacs,nofootinbib,preprintnumbers]{revtex4}
\usepackage{amssymb,amsmath,epsfig}

\begin{document}

\title{Exclusive production of quarkonia as a probe of the GPD {\boldmath $E$} for gluons}

\author{John Koempel}
\affiliation{Department of Physics, Barton Hall,
             Temple University, Philadelphia, PA 19122, USA}

\author{Peter Kroll}
\affiliation{Fachbereich Physik, Universit\"at Wuppertal,
             42097 Wuppertal, Germany}
\affiliation{Institut f\"ur Theoretische Physik,
             Universit\"at Regensburg, 93040 Regensburg, Germany}

\author{Andreas Metz}
\affiliation{Department of Physics, Barton Hall,
             Temple University, Philadelphia, PA 19122, USA}


\author{Jian Zhou}
\affiliation{Institut f\"ur Theoretische Physik,
             Universit\"at Regensburg, 93040 Regensburg, Germany}

\begin{abstract}
Exclusive quarkonium photo- and electro-production off the nucleon is studied in the 
framework of generalized parton distributions (GPDs).
The short distance part of the process is treated at leading order in perturbative
Quantum Chromodynamics.
The main focus is on the GPD $E$ for gluons.
On the basis of different models for $E^g$ we estimate the transverse target spin 
asymmetry for typical kinematics of a future Electron Ion Collider.
We also explore the potential of measuring the polarization of the recoil nucleon.
\end{abstract}

\pacs{12.38.-t; 12.39.St; 13.60.Le; 13.88.+e}

\date{\today}

\maketitle

\noindent
I.~{\it Introduction.}\,---\,For about 15 years, 
GPDs~\cite{Mueller:1998fv,Ji:1996ek,Radyushkin:1996nd,Radyushkin:1996ru,Ji:1996nm,
Goeke:2001tz,Diehl:2003ny,Belitsky:2005qn,Boffi:2007yc} have been playing a key role 
in hadronic physics for a number of reasons.
First, GPDs serve as unifying objects, containing the information encoded both in 
ordinary parton distributions and in form factors.
Second, GPDs allow us to explore the partonic structure of hadrons in three 
dimensions~\cite{Burkardt:2000za,Ralston:2001xs,Diehl:2002he,Burkardt:2002hr}.
Third, GPDs enter Ji's spin sum rule of the nucleon~\cite{Ji:1996ek}.

GPDs can be measured in hard exclusive processes like deep-virtual Compton 
scattering off the nucleon or hard exclusive meson 
production~\cite{Ji:1996ek,Radyushkin:1996nd,Radyushkin:1996ru,Ji:1996nm,
Collins:1996fb,Collins:1998be}. 
They depend on three kinematical variables: the average momentum fraction 
$x$ of the partons, the longitudinal momentum transfer $\xi$ to the nucleon 
(skewness), and the invariant momentum transfer $t$ of the process, i.e.,
$F = F(x,\xi,t)$ for a generic GPD $F$.
According to~\cite{Ji:1996ek}, GPDs give access to the angular momenta of 
quarks and gluons inside the nucleon, where the total spin of the nucleon is 
given by $\frac{1}{2} = \sum_{q} J^q + J^g$.
Specifically, the gluon angular momentum can be determined 
through~\cite{Ji:1996ek}
\begin{equation}
J^g = \frac{1}{2} \int_0^1 dx 
\Big( H^g(x, \xi, 0) + E^g(x, \xi, 0) \Big) ,
\label{e:J_gluon}
\end{equation}
with $H^g$ and $E^g$ denoting the dominant (leading twist) GPDs of unpolarized
gluons inside the nucleon.
Considerable information on $H^g$ is already available, for it is connected to 
the ordinary unpolarized gluon distribution via $H^g(x,0,0) = x g(x)$ --- see for 
instance Refs.~\cite{Frankfurt:2005mc,Goloskokov:2005sd,
Goloskokov:2006hr,Diehl:2007hd,Goloskokov:2007nt,Martin:2009zzb,Kumericki:2009uq}.
In comparison, our knowledge about $E^g$ is still marginal.
Therefore, in particular, the value for $J^g$ in Eq.~(\ref{e:J_gluon}) is
still very uncertain.

It has been known for quite some time that exclusive quarkonium ($J/\psi$ or $\Upsilon$)
production off the nucleon is very suitable for probing the gluonic structure of 
the nucleon in a clean way~\cite{Ryskin:1992ui,Brodsky:1994kf}, since quark 
exchange plays only a minor role.
Moreover, due to the large scale provided by the heavy quark/meson mass, perturbative
Quantum Chromodynamics (QCD) can be applied even for photo-production.
In the present work, we consider both photo- and electro-production of $J/\psi$ and 
$\Upsilon$ off a proton target.
Using leading order (LO) results for the hard scattering coefficients we study the 
prospects for measuring $E^g$ by means of quarkonium production.
To this end we consider several models for $E^g$, and compute the transverse target 
spin asymmetry as well as a double spin observable which requires polarimetry of 
the recoil nucleon.
We provide numerical results for typical kinematics of a potential future Electron 
Ion Collider (EIC)~\cite{Anselmino:2011ay,Boer:2011fh,Accardi:2011mz}.

\noindent
II.~{\it Theoretical framework.}\,---\,For definiteness, we consider the 
process
\begin{equation} 
\gamma^{(\ast)}(q,\mu) + p(p,\nu) \to V(q',\mu') + p(p',\nu') ,
\label{e:process}
\end{equation}
where the 4-momenta and the helicities of the particles are specified.
We further use $Q^2 = - q^2$, $m^2 = p^2 = p'^2$, $m_V^2 = q'^2$, $t = (p - p')^2$,
and the squared photon-nucleon {\it cm} energy $W^2 = (p + q)^2$.
The skewness variable can be expressed as
\begin{equation}
\xi = \frac{\tilde{x}_{B}}{2 - \tilde{x}_{B}} \,, \quad \textrm{with} \quad
\tilde{x}_B = \frac{m_V^2 + Q^2}{W^2 + Q^2} \,, 
\end{equation}
which holds for arbitrary values of $Q^2$.
The minimal value of $t$ is given by $|t_0| = 4 m^2 \xi^2 / (1-\xi^2)$.

For $Q^2$ much larger than all other scales of the process, an all order proof of 
QCD factorization has been formulated for the process 
in~(\ref{e:process})~\cite{Collins:1996fb}.
In the case of quarkonium production, one may expect factorization to hold for 
arbitrary $Q^2$.
In fact, a next-to-leading order (NLO) calculation of the unpolarized photo-production 
cross section is compatible with factorization~\cite{Ivanov:2004vd}.
\begin{figure}[t]
\begin{center}
\includegraphics[width=5.0cm]{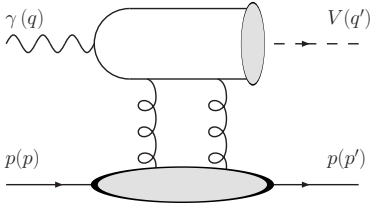}
\caption{Sample LO diagram for the process in~(\ref{e:process}).
The lower part of the diagram is parameterized in terms of gluon GPDs.}
\label{f:sample}
\end{center}
\end{figure}

To LO in the strong coupling there are six Feynman graphs --- see  
Fig.$\,$\ref{f:sample} for a sample diagram.
They factorize into the hard subprocess $\gamma^{\ast} g \to V g$ and gluon GPDs.
We computed the helicity amplitudes of the subprocess in the non-relativistic 
approximation for which the heavy quark and antiquark carry the same momentum. 
Our results agree with previous 
calculations~\cite{Vanttinen:1998en,Ivanov:2004vd,Koempel:prep}.
The structure of the LO amplitudes implies that one is sensitive only to the 
GPDs $H^g$ and $E^g$~\cite{Vanttinen:1998en,Koempel:prep}.

For transversely polarized photons and vector mesons, the nonzero helicity 
amplitudes ${\cal M}_{\mu' \nu' , \mu \nu}$ of the full process read
\begin{eqnarray}
{\cal M}_{\pm + , \pm +} & = & {\cal M}_{\pm - , \pm -} = 
C \sqrt{1 - \xi^2} \, \langle H_{\rm eff}^g \rangle ,
\phantom{\frac{1}{2}}
\label{e:nf_t}
\\
{\cal M}_{\pm - , \pm +} & = & - {\cal M}_{\pm + , \pm -} = 
- C \frac{\sqrt{-t'}}{2m} \, \langle E^g \rangle ,
\label{e:sf_t}
\\
\textrm{with} & & \hspace{-0.5cm}
\langle F \rangle = \int_0^1 \frac{dx}{(x + \xi)(x - \xi + i\varepsilon)} \,
F(x,\xi,t) 
\nonumber
\end{eqnarray}
for a generic GPD $F$.
In Eqs.~(\ref{e:nf_t}),~(\ref{e:sf_t}) we use the definitions 
$H_{\rm eff}^g = H^g - \xi^2/(1-\xi^2) E^g$, $t' = t - t_0$, 
and $C = 16 \pi e_q e \alpha_s f_V m_V / (N_c (Q^2 + m_V^2))$, where
$f_V$ denotes the quarkonium decay constant. 
Moreover, one has
\begin{equation}
{\cal M}_{0 \nu' , 0 \nu} = - \frac{Q}{m_V} {\cal M}_{\pm \nu' , \pm \nu} 
\end{equation}
for longitudinal transitions.
A corresponding relation between the longitudinal and the transverse amplitudes 
was previously obtained in the pioneering work on exclusive $J/\psi$ production 
in the leading double-log approximation~\cite{Ryskin:1992ui}.

\noindent
III.~{\it Generalized Parton Distributions.}\,---\,For the GPD $H$ we take the 
parameterizations obtained in previous 
analyses~\cite{Goloskokov:2006hr,Goloskokov:2007nt}.
Note that quark GPDs are also needed since the GPDs are evolved to different 
scales.
In the case of $E$, we use the valence quark distributions from 
Ref.~\cite{Diehl:2004cx,Goloskokov:2008ib}, while we explore different scenarios for gluons 
and seaquarks.
They are modelled through double 
distributions~\cite{Mueller:1998fv,Musatov:1999xp} according to
\begin{eqnarray}
E^{i}(x,\xi,t) & \!=\! & \int_{-1}^{1} d\beta \int_{-1+|\beta|}^{1-|\beta|} d\alpha \,
\delta(\beta + \xi\alpha - x) f^{i}(\beta,\alpha,t) ,
\nonumber \\
f^{i}(\beta,\alpha,t) & \!= \! & E^{i}(\beta,0,t)
\frac{15}{16} \frac{[(1 - |\beta|)^2 - \alpha^2]^2}{(1 - |\beta|)^{5}} ,
\\
\textrm{with} & & \hspace{-0.2cm} 
E^i(\beta,0,t) = e^{b_e t} |\beta|^{-\alpha'_e t} E^i(|\beta|,0,0) .
\nonumber
\end{eqnarray}
For gluons we define $x e^g(x) \equiv E^g(x,0,0)$ and investigate 
the two cases
\begin{eqnarray}
e^{g}(x) & = & N^{g} x^{-1-\delta_e} (1 - x)^{\beta_e^g}, 
\label{e:ansatz_1} \\
e^{g}(x) & = & N^{g} x^{-1-\delta_e} (1 - x)^{\beta_e^g} \tanh (1 - x/x_0),
\label{e:ansatz_2}
\end{eqnarray}
where the ansatz in~(\ref{e:ansatz_2}) has a node at $x = x_0$.
Such a possibility is currently not ruled out~\cite{Diehl:INT_talk}.
We further define $e^q(x) \equiv E^q(x,0,0)$, and use a flavor-symmetric sea, i.e., 
$e^{\bar{q}} \equiv e^{\bar{u}} = e^{\bar{d}} = e^{\bar{s}} = e^{s}$.
For $e^{\bar{q}}(x)$ we make an ansatz analogous to~(\ref{e:ansatz_1}) and do not 
consider a node.
For the parameters $b_e$, $\alpha_e^{\prime}$, and $\delta_e$ we do not distinguish 
between gluons and seaquarks.
\begin{table}[t]
\renewcommand{\arraystretch}{1.2} 
\begin{center}
\begin{tabular}{|c|c||c|c|c|c||c||c|}
\hline  
 Var. & $\alpha_e'$ & $N^g$ & $x_0$ & $e_{20}^g$ & $J^g$ &  
 $N^{\bar{q}}$ & $J^s$ \\
\hline
1 & $\hspace{0.05cm} 0.15 \hspace{0.05cm}$  & 0 & & 0 & $\hspace{0.05cm} 0.214 \hspace{0.05cm}$ 
    & $\hspace{0.05cm}-0.009\hspace{0.05cm}$ & $\hspace{0.05cm} 0.014 \hspace{0.05cm}$ \\ 
2 & 0.15 & $\hspace{0.05cm}-0.878\hspace{0.05cm}$ & & $\hspace{0.05cm}-0.164\hspace{0.05cm}$ 
    & 0.132 & $\phantom{-}0.156$ & 0.041 \\ 
3 & 0.10 & $-1.017$ & & $-0.190$ & 0.119 & $\phantom{-}0.182$ & 0.045 \\
4 & 0.10 & $\phantom{-}3.015$ & 0.05 & $-0.190$ & 0.119 & $\phantom{-}0.182$ & 0.045 \\
5 & 0.10 & $-1.974$ & $\hspace{0.05cm}0.3\phantom{0}\hspace{0.05cm}$ & $-0.190$ & 0.119 
    & $\phantom{-}0.182$ & 0.045 \\ 
\hline
\end{tabular}
\end{center}
\caption{Parameters of $e^g$ and $e^{\bar{q}}$ at the scale $\mu = 2\,\textrm{GeV}$. 
For gluons, the Variants 1,2,3 refer to the ansatz in~(\ref{e:ansatz_1}), while 
Variants 4,5 refer to~(\ref{e:ansatz_2}).
Also shown is the second moment $e_{20}^g$, and values for the angular momenta as
defined in~(\ref{e:J_gluon}).}
\label{tab:e}
\renewcommand{\arraystretch}{1.0}   
\end{table} 

Two constraints have to be satisfied when fixing the parameters for $e^{g}$ and $e^{\bar{q}}$.
First, the momentum sum rule for unpolarized parton distributions in combination
with Ji's spin sum rule leads to
\begin{eqnarray}
e_{20}^g & \!=\! & - \sum_q e_{20}^{q_{val}} - 2 \sum_q e_{20}^{\bar{q}} , \quad
\textrm{with}
\label{e:mom_rel}
\\
e_{n0}^i & \!\equiv\! & \int_0^1 dx \, x^{n-1} e^i(x) .
\end{eqnarray}
Second, the density interpretation of GPDs in the impact parameter 
space~\cite{Burkardt:2002hr} leads to a positivity bound for $e^g$ and 
$e^{\bar{q}}$ --- see Refs.~\cite{Diehl:2004cx,Diehl:2007hd,Goloskokov:2008ib,Koempel:prep} 
for more details.
In particular, one finds $b_e < b_h$, $\alpha_e^{\prime} \le \alpha_h^{\prime}$,
where $b_h$ and $\alpha_h^{\prime}$ appear in the double distribution ansatz of 
$H^g$.
We take 
$b_h = 2.58 \, \textrm{GeV}^{-2} + 0.25 \, \textrm{GeV}^{-2} \ln \frac{m^2}{m^2 + \mu^2}$ 
(with $\mu$ representing the scale of the GPD) and $\alpha_h^{\prime} = 0.15$ from 
previous work~\cite{Goloskokov:2006hr,Goloskokov:2007nt}.
We choose $b_e = 0.9 \, b_h$ and explore two different values for $\alpha_e^{\prime}$
(see Tab.$\,$\ref{tab:e}).
Moreover, we use $\delta^e = 0.1$, as well as $\beta_e^g = 6$ and 
$\beta_e^{\bar{q}} = 7$~\cite{Goloskokov:2008ib}.
After these choices have been made, only the normalization constants $N^g$ and 
$N^{\bar{q}}$ remain to be determined.

\begin{figure}[t]
\begin{center}
\includegraphics[width=0.22\textwidth]{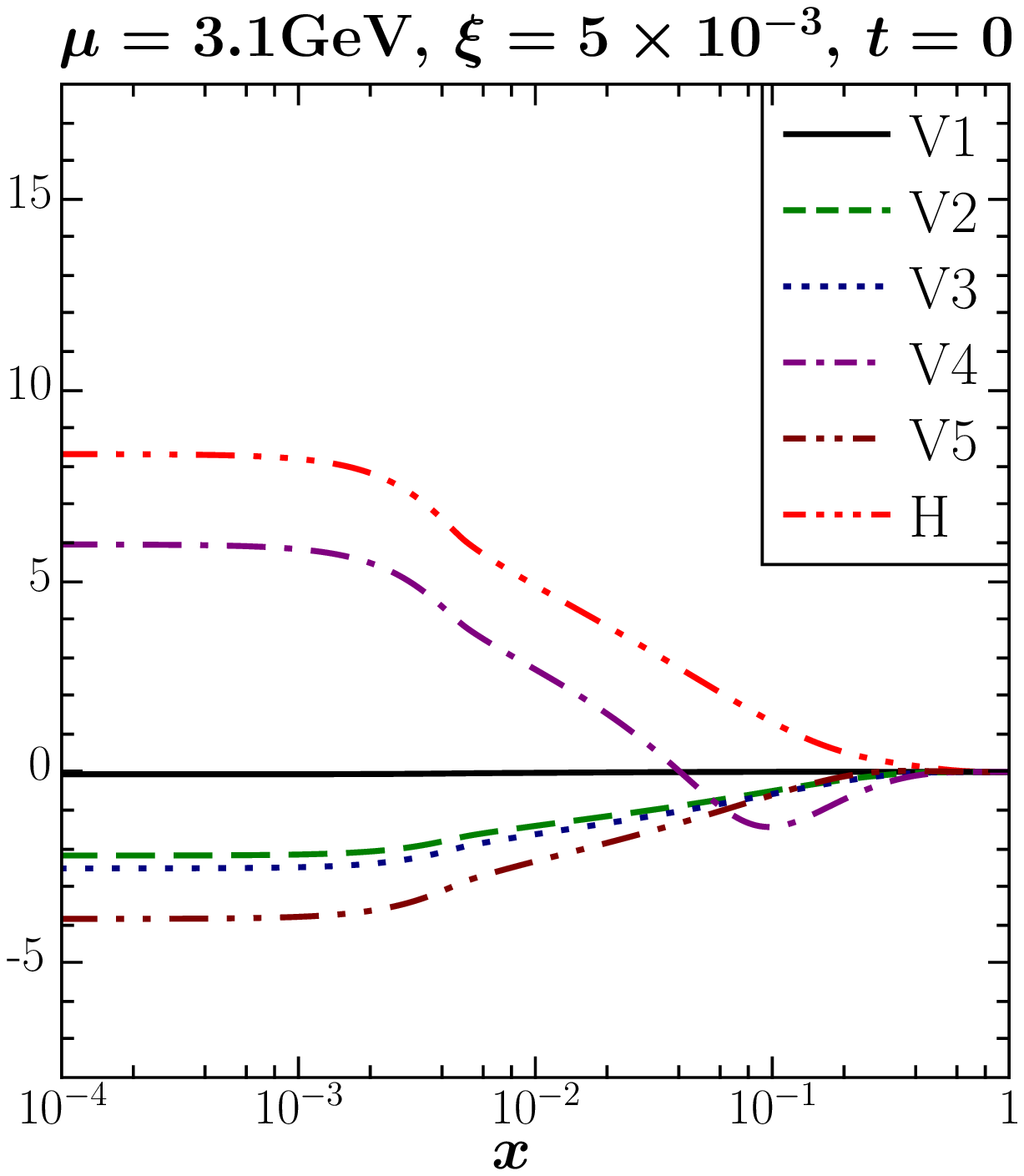}
\hskip 0.3cm
\includegraphics[width=0.22\textwidth]{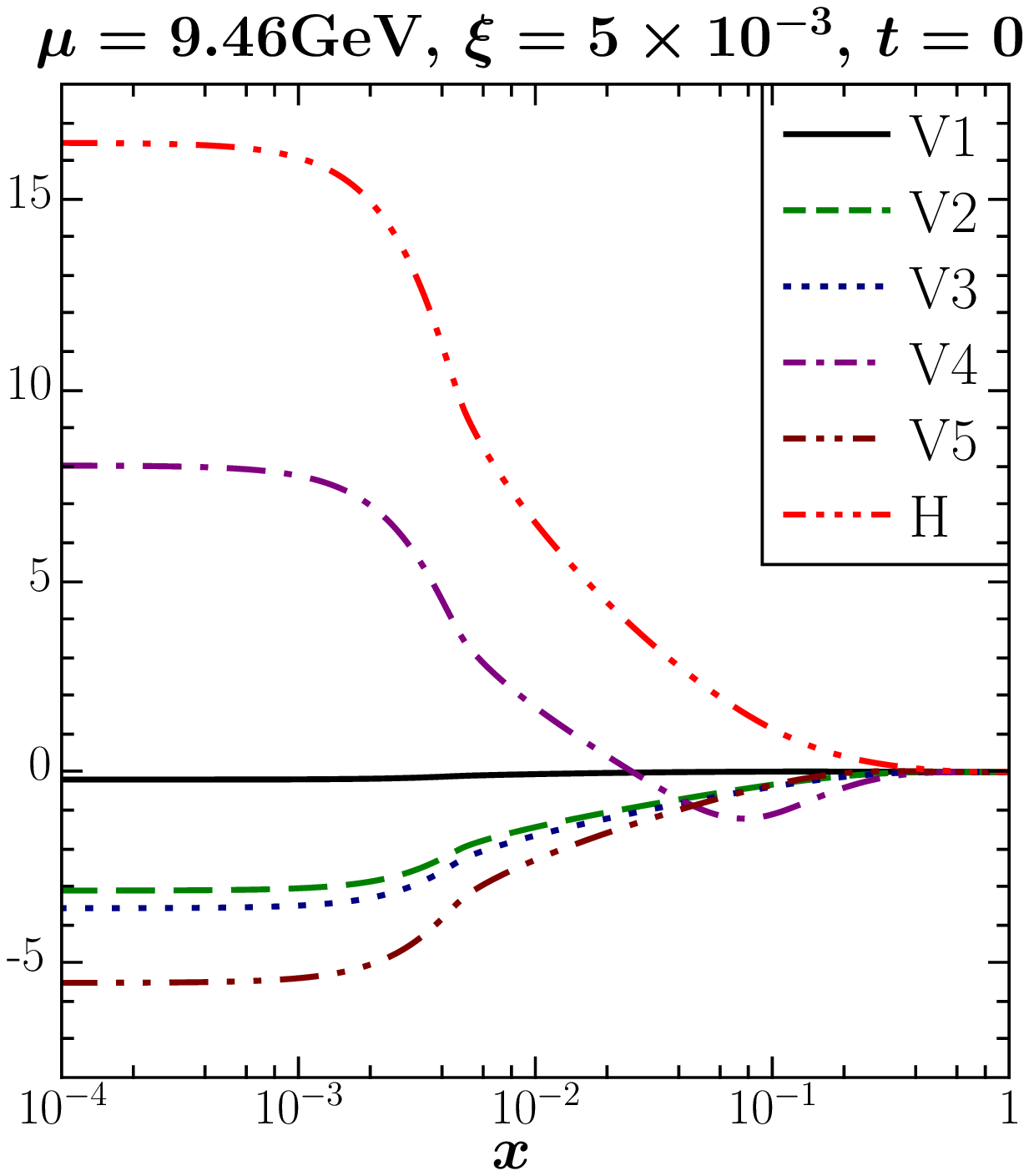}
\caption{Variants 1--5 for $E^g$, together with $H^g$, 
at the scales $\mu = m_{J/\psi} = 3.1 \, \textrm{GeV}$ (left) 
and $\mu = m_{\Upsilon} = 9.46 \, \textrm{GeV}$ (right).} 
\label{f:GPDs}
\end{center}
\end{figure}

We parameterize $e^{g}$ and $e^{\bar{q}}$ at the scale $\mu = 2 \, \textrm{GeV}$.
For $e^g$ we consider five different variants, where the respective parameters are 
listed in Tab.$\,$\ref{tab:e}.
Variant 1 means $e^g = 0$, and the normalization $N^{\bar{q}}$ is fixed by means of
the relation in~(\ref{e:mom_rel}).
(There is actually some support for a rather small $E^g$: this GPD has a 
model-dependent relation to the transverse momentum dependent gluon Sivers 
function~\cite{Meissner:2007rx}, which may be 
small~\cite{Efremov:2004tp,Anselmino:2006yq,Brodsky:2006ha}.) 
In the remaining four cases we first determine $N^{\bar{q}}$ by saturating the 
positivity bound, then compute $N^{g}$ from~(\ref{e:mom_rel}), and finally check
whether $e^g$ satisfies the positivity bound.
Variants 4 and 5 for $e^g$ contain a node.
The positivity bound does not allow one to fix the sign of $N^{\bar{q}}$.
However, none of our general conclusions depends on this sign~\cite{Koempel:prep}.
We also checked that all variants for $e^{g}$ and $e^{\bar{q}}$ are compatible with 
a preliminary data point for the transverse target spin asymmetry for exclusive 
$\phi$ production from HERMES~\cite{Augustiniak:DIS_talk,Koempel:prep}.
When calculating observables we evolve the GPDs to the scale $\mu = (m_V^2 + Q^2)^{1/2}$
by using the code of Vinnikov~\cite{Vinnikov:2006xw}.
In Fig.$\,$\ref{f:GPDs}, the GPDs are displayed at the scale of the quarkonium masses.
\begin{figure}[!]
\begin{center}
\includegraphics[width=0.22\textwidth]{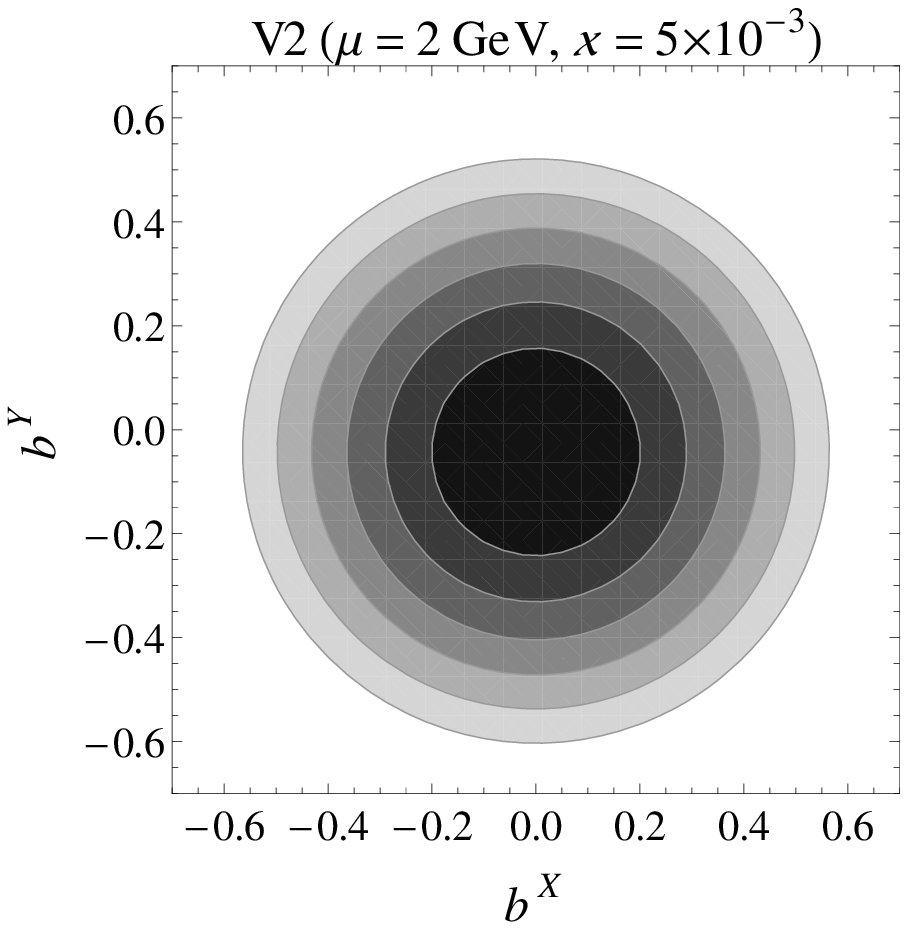}
\hskip 0.3cm
\includegraphics[width=0.22\textwidth]{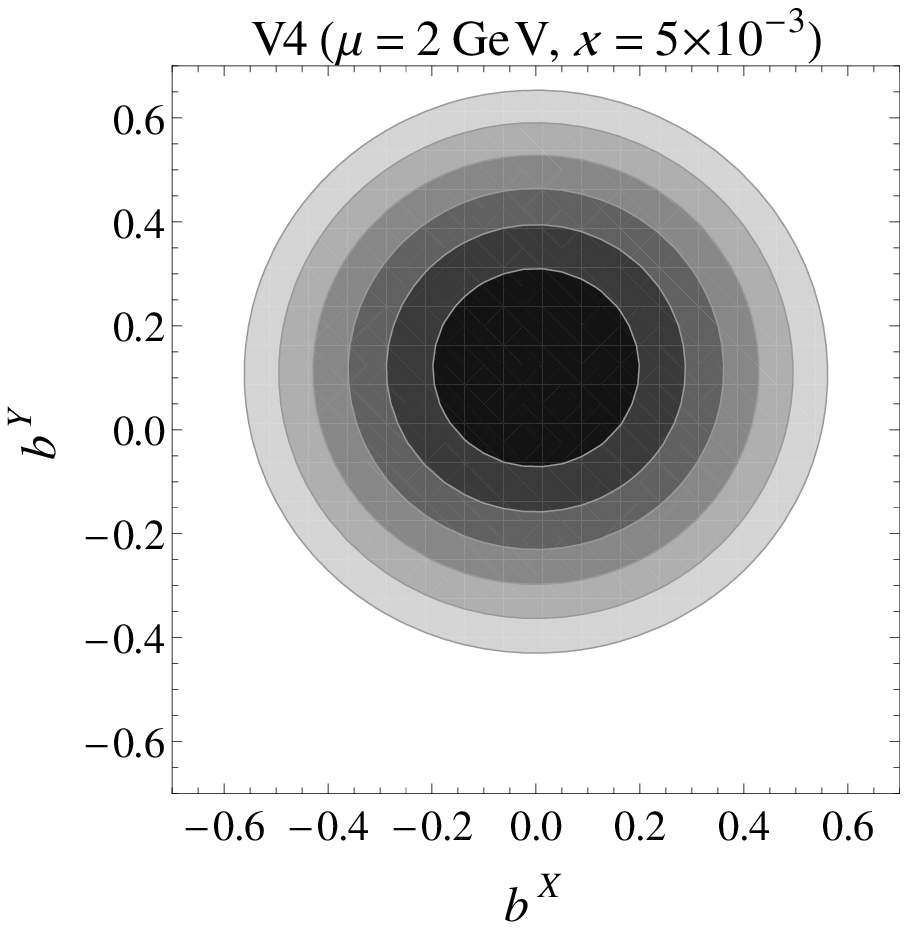}
\caption{Density in~(\ref{e:density}) for Variant 2 (left) and Variant 4 (right) at 
$x = 0.005$ and $\mu = 2 \, \textrm{GeV}$.
The outer ring indicates half of the maximum density.}
\label{f:tomography}
\end{center}
\end{figure}

Following~\cite{Ji:1996ek}, the gluon and strange quark contributions to the
nucleon spin are $J^g =  (h_{20}^g + e_{20}^g)/2$ as well as 
$J^{s} =  (h_{20}^s + e_{20}^s)$,
where the densities $h^{g/q}$ are related to $H^{g/q}$ in the same way as $e^{g/q}$ 
are related to $E^{g/q}$.
Taking $h_{20}^g = 0.4276$ and $h_{20}^s = 0.0153$ from~\cite{Pumplin:2002vw,Goloskokov:2008ib} 
leads to the values for $J^g$ and $J^{s}$ in Tab.$\,$\ref{tab:e}.
Because of~(\ref{e:mom_rel}), a change of $J^{s}$ implies a change of $J^g$.
For our parameterizations, the contribution from $E^g$ to the nucleon spin can be 
significant (up to about $20\,\%$).

One can also calculate the density of unpolarized gluons in transverse position 
(impact parameter $b_{\perp}$) space.
If the nucleon is transversely polarized (along the $X$-direction), this density 
is given by~\cite{Burkardt:2002hr}
\begin{equation}
\mathcal{H}^{g,X}(x,\vec{b}_{\perp}) = \mathcal{H}^{g}(x,\vec{b}_{\perp}^{\,2}) 
 - \frac{b_{\perp}^{Y}}{m} \frac{\partial}{\partial \vec{b}_{\perp}^{\,2}} \, 
 \mathcal{E}^{g}(x,\vec{b}_{\perp}^{\,2}) ,
\label{e:density}
\end{equation}
with $\mathcal{H}^g$ and $\mathcal{E}^g$ denoting the $b_{\perp}$-space representation 
of $H^g$ and $E^g$, respectively.
The sample plots in Fig.$\,$\ref{f:tomography} show, in particular, how much the maximum 
of the density in~(\ref{e:density}) is shifted away from the origin due to the 
presence of the $E^g$-term.
\begin{figure}[t]
\begin{center}
\includegraphics[width=0.22\textwidth]{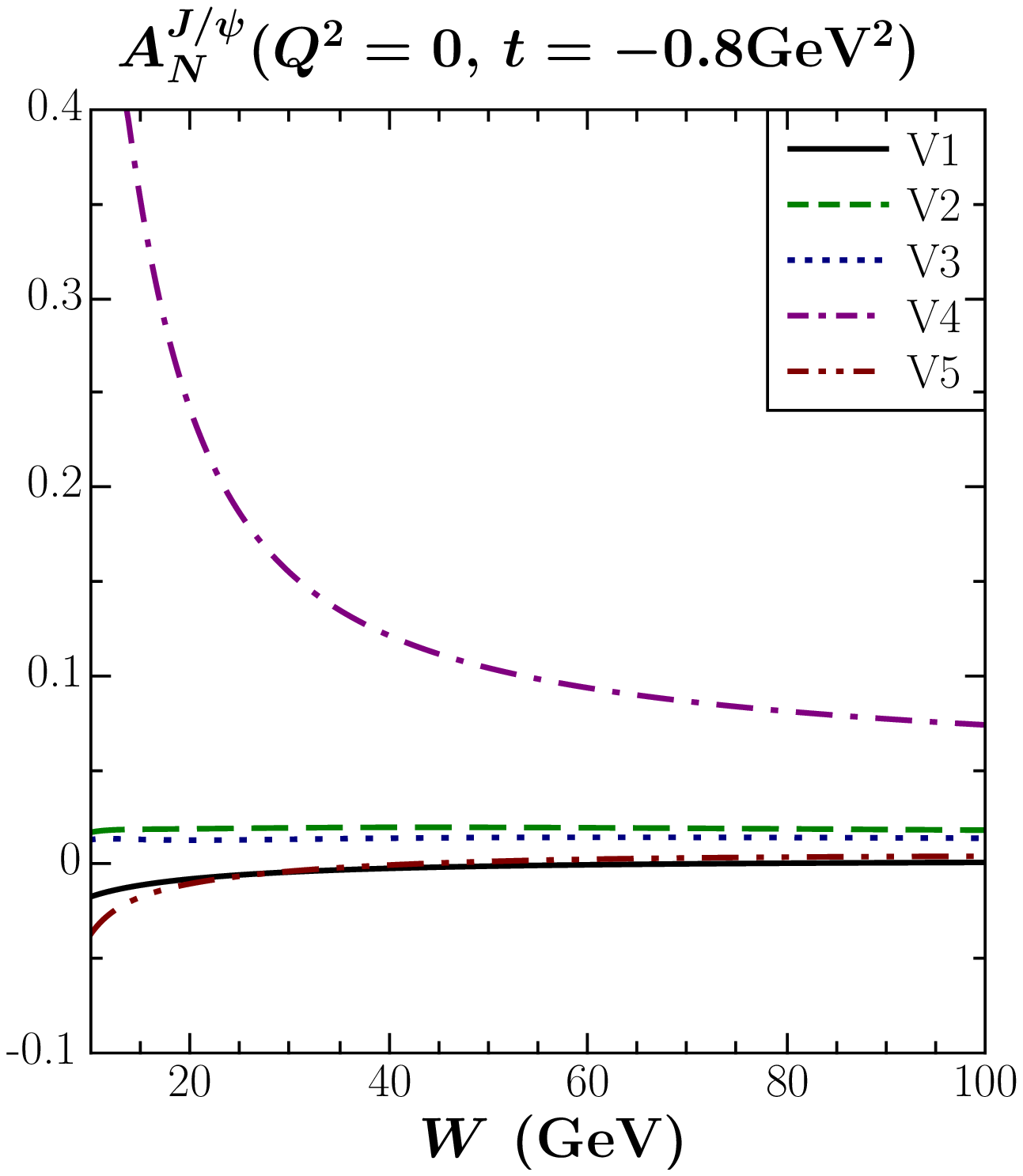}
\hskip 0.3cm
\includegraphics[width=0.22\textwidth]{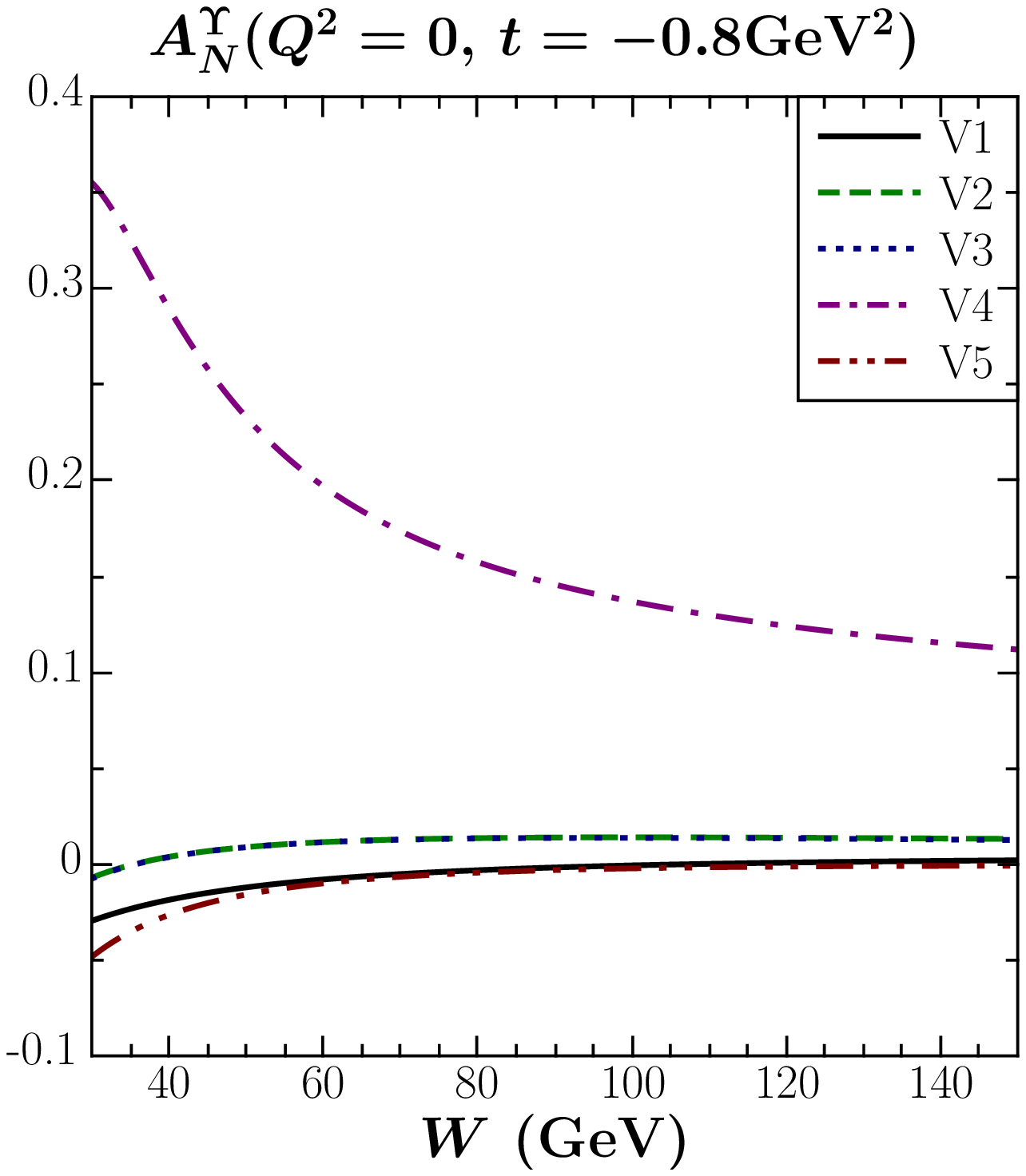}
\caption{SSA $A_{N}$ in~(\ref{e:ssa}) for photo-production of $J/\psi$ (left) and 
$\Upsilon$ (right) as function of $W$ for different variants of $E^g$.}
\label{f:AUT_W}
\end{center}
\end{figure}

\noindent
IV.~{\it Polarization observables.}\,---\,As discussed in Sect.~II, one has two 
independent amplitudes: the non-flip amplitude ${\cal M_{++,++}}$, dominated
by $H^g$, and the spin-flip amplitude ${\cal M_{+-,++}}$, which is determined by 
$E^g$. 
The following observables allow one to measure those amplitudes (up to an 
overall phase): the unpolarized cross section, the transverse target single spin 
asymmetry (SSA) $A_{N}$ (with the polarization being normal to the reaction plane --- 
often $A_N$ is also denoted as $A_{UT}$), and two double spin asymmetries (DSAs) 
requiring a polarized target and polarimetry of the recoil nucleon~\cite{Koempel:prep}.
Here we focus on
\begin{eqnarray}
A_{N} & = & \frac{- 2 \, {\rm Im}({\cal M}_{++,++} \, {\cal M}_{+-,++}^{\ast})}
{|{\cal M}_{++,++}|^2 + |{\cal M}_{+-,++}|^2} ,
\label{e:ssa}
\\
A_{LS} & = & \frac{2 \, {\rm Re}({\cal M}_{++,++} \, {\cal M}_{+-,++}^{\ast})}
{|{\cal M}_{++,++}|^2 + |{\cal M}_{+-,++}|^2} ,
\label{e:dsa}
\end{eqnarray}
where for the DSA $A_{LS}$ the target is longitudinally polarized, and the 
outgoing nucleon is transversely polarized 
(in the reaction plane, ``sideways'')~\cite{Koempel:prep}.
We consider production of both $J/\psi$ and $\Upsilon$ for typical EIC kinematics.
While the $J/\psi$ final state has a larger cross section, one can expect a better 
convergence of the $\alpha_s$-expansion in the case of the $\Upsilon$~\cite{Ivanov:2004vd}.
A detailed comparison with existing data for the unpolarized cross section will be given 
elsewhere~\cite{Koempel:prep} --- see also Ref.~\cite{Ivanov:2004vd}. 
\begin{figure}[t]
\begin{center}
\includegraphics[width=0.22\textwidth]{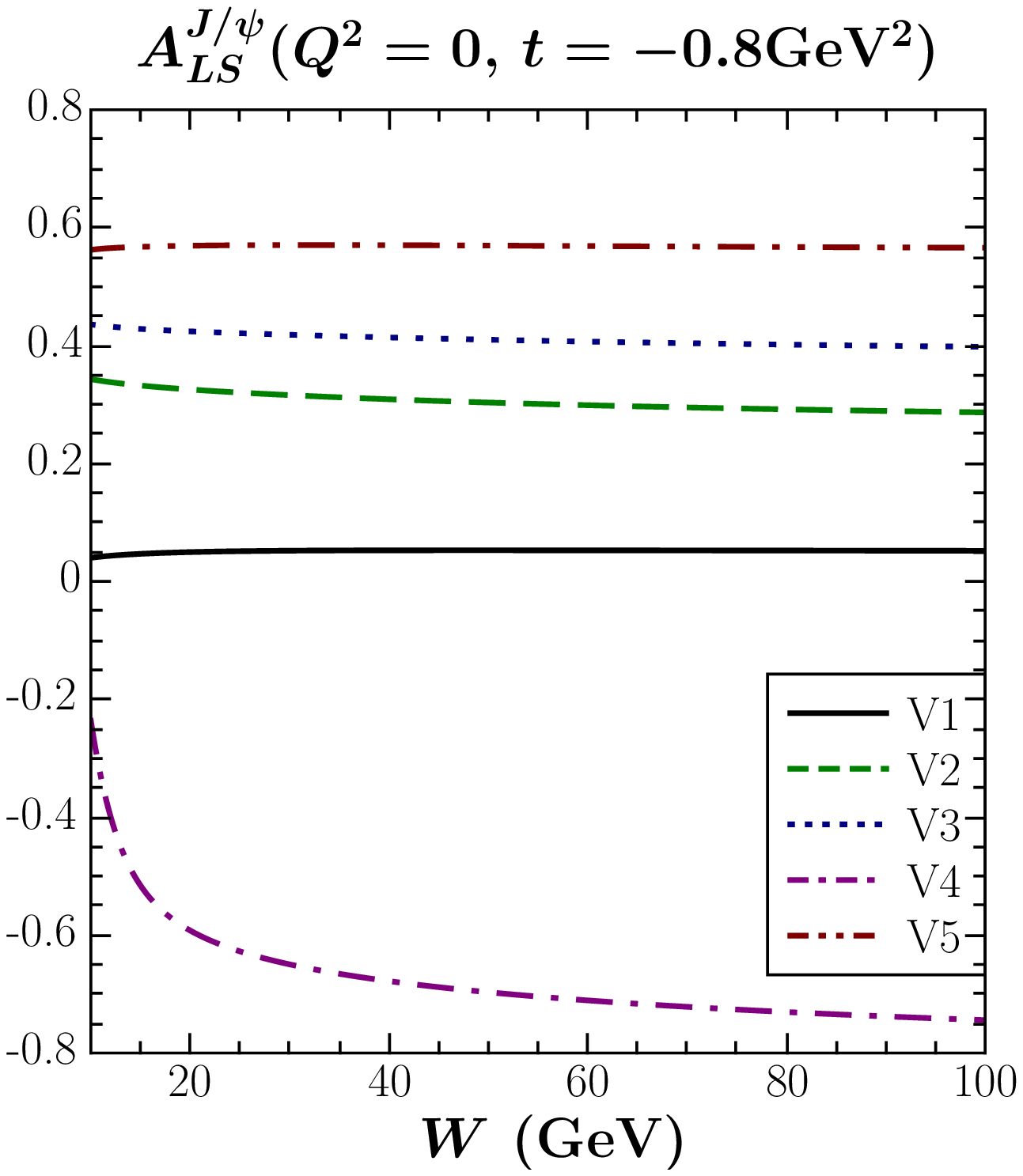}
\hskip 0.3cm
\includegraphics[width=0.22\textwidth]{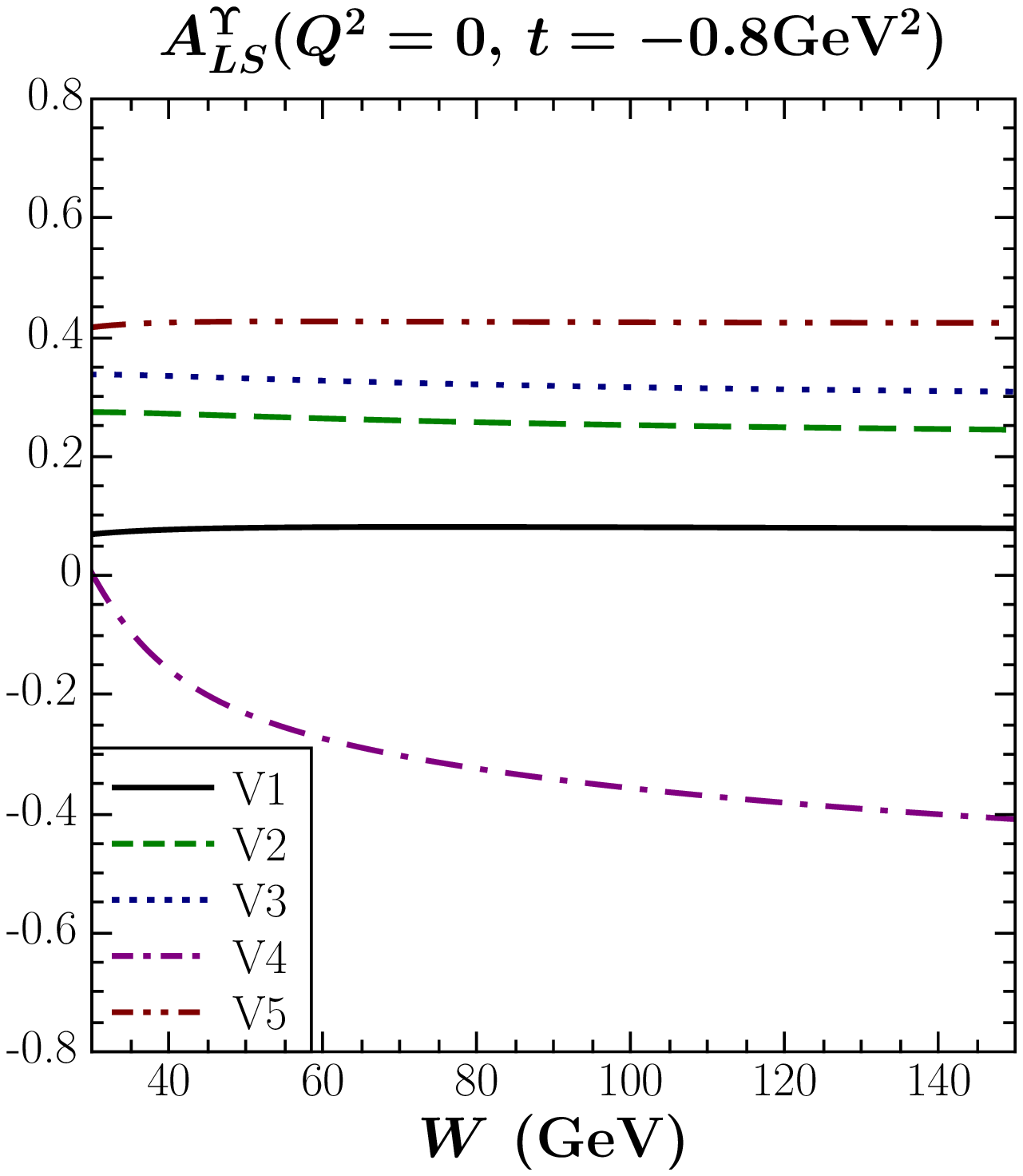}
\caption{DSA $A_{LT}$ in~(\ref{e:dsa}) for photo-production of $J/\psi$ (left) and 
$\Upsilon$ (right) as function of $W$ for different variants of $E^g$.}
\label{f:ALT_W}
\end{center}
\end{figure}
\begin{figure}[!]
\begin{center}
\includegraphics[width=0.22\textwidth]{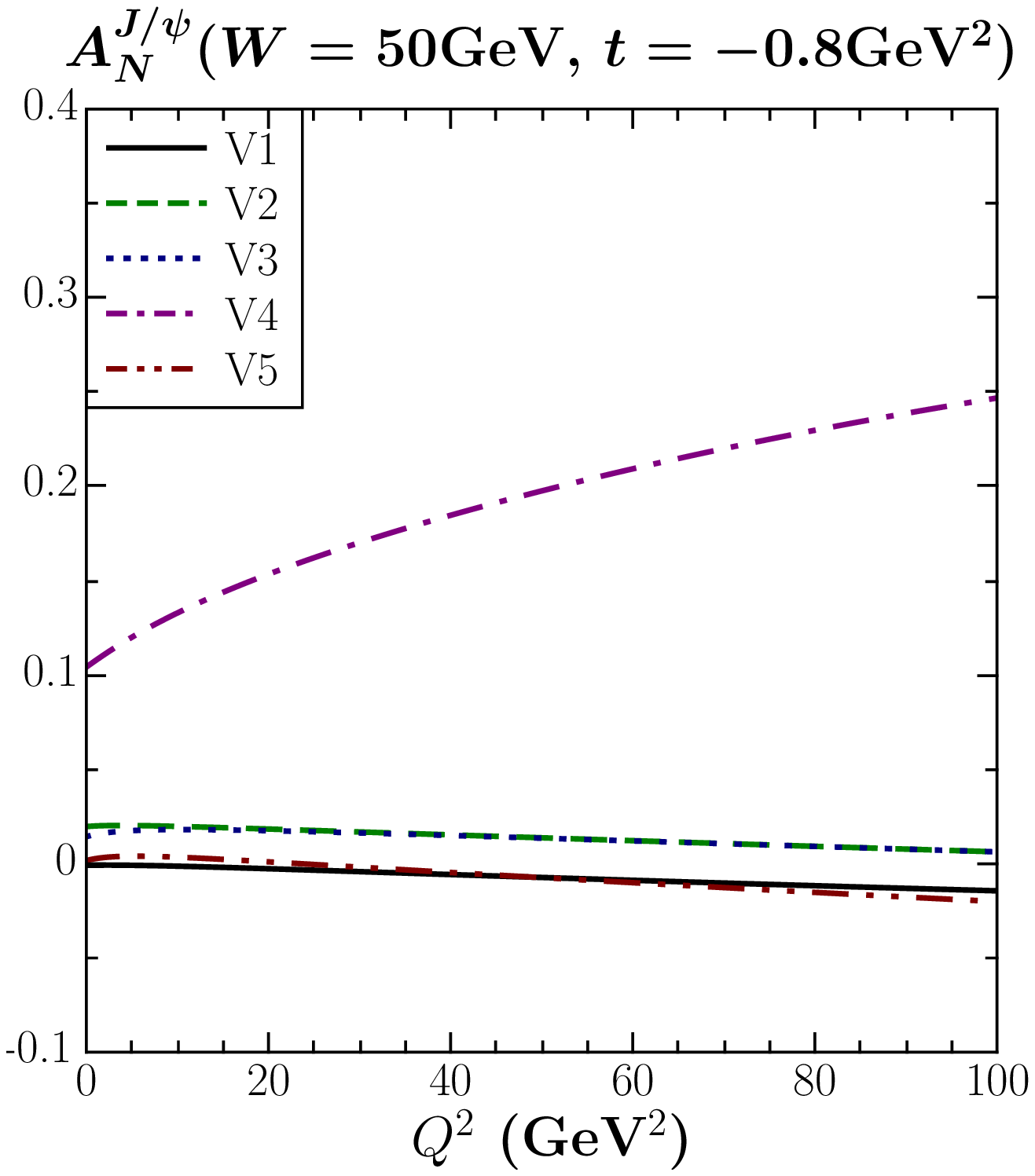}
\hskip 0.3cm
\includegraphics[width=0.22\textwidth]{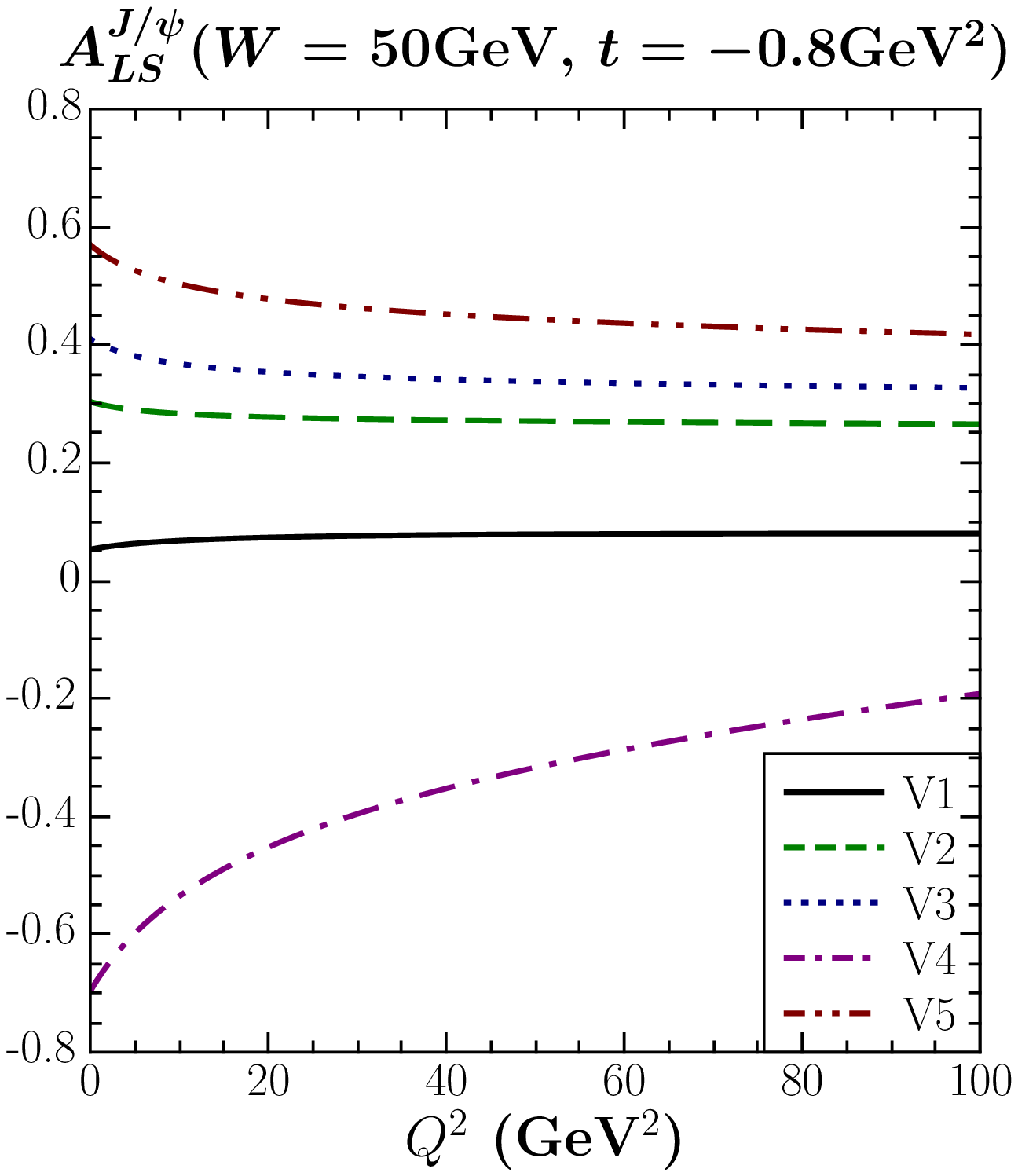}
\caption{$A_{N}$ (left) and $A_{LS}$ (right) for production of $J/\psi$
as function of $Q^2$ for different variants of $E^g$.}
\label{f:asymm_Q}
\end{center}
\end{figure}

In Fig.$\,$\ref{f:AUT_W}, $A_N$ is shown for photo-production of $J/\psi$ and $\Upsilon$
as a function of $W$.
This asymmetry is rather small for most variants of $E^g$, mainly because the
respective non-flip amplitude and the spin-flip amplitude have a similar phase.
It can also be seen, however, that larger values of $A_N$ are currently not ruled out.
On the basis of a LO calculation one can not draw a definite conclusion about the optimal
$W$ for a measurement of $A_N$.
But higher order terms to the unpolarized cross section are better under control 
for lower values of $W$~\cite{Ivanov:2004vd}. 
In general, the spin asymmetries are less influenced by NLO corrections than the cross 
section~\cite{Koempel:prep}.
The DSA $A_{LS}$ is displayed in Fig.$\,$\ref{f:ALT_W}.
This observable is small only if $E^g$ is small.
It is worthwhile to explore the feasibility of a corresponding measurement, since $A_{LS}$ 
may give the most direct access to $E^g$.
(We note that, from a theoretical point of view, the DSA $A_{SL}$ is equally well 
suited~\cite{Koempel:prep}.)
Finally, the $Q^2$-dependence of $A_{N}$ and $A_{LS}$ for $J/\psi$ production is shown
in Fig.$\,$\ref{f:asymm_Q}.
Electro-production at low $Q^2$ is attractive because of the large count rates.
For $Q^2 \ge m_V^2$, higher order corrections may be less important~\cite{Ivanov:2004vd}, 
but an explicit calculation does not yet exist.

\noindent
V.~{\it Summary}\,---\,We have explored the potential of measuring the GPD $E^g$
through exclusive production of quarkonium at a future Electron Ion Collider. 
The study is based on a LO calculation of the short distance part of the process,
and several models for $E^g$ which respect the currently known constraints.
Most variants of $E^g$ lead to a rather small transverse target SSA $A_N$, but
a healthy $A_N$ is presently not ruled out either.
We have also found promising results for a double polarization observable (polarized 
target and polarized recoil nucleon), which provides a quite direct access to $E^g$.

\begin{acknowledgments}
We thank Markus Diehl, Zein-Eddine Meziani, and Feng Yuan for useful discussions and 
encouragement.
This work was supported in part by the BMBF under the contract No.~06RY258, and the
NSF under the Grant No.~PHY-0855501.
\end{acknowledgments}


\end{document}